\documentclass{elsarticle}

\usepackage[T1]{fontenc}
\usepackage{amssymb}
\usepackage{geometry}
\usepackage[hidelinks]{hyperref}
\usepackage{multirow}
\usepackage{subcaption}
\usepackage{verbatim}

\graphicspath{{./figures/}}
\newcommand{\h}[1]{\rotatebox{90}{#1\hspace{1ex}}}
\newcommand{\y}{$\blacksquare$}
\newcommand{\n}{$\square$}

\journal{Computer Languages, Systems \& Structures}
\usepackage[absolute]{textpos}
\makeatletter
\def\ps@pprintTitle{
\def\@oddfoot{}
}
\makeatother

\begin{document}

\begin{textblock}{15}(0.5,14.5)
{
\vspace{-2pt}\noindent\hrulefill

\noindent\fontsize{8pt}{8pt}\selectfont\copyright\ 2018. This manuscript version is made available under the CC-BY-NC-ND 4.0 license: \url{http://creativecommons.org/licenses/by-nc-nd/4.0/}. \hspace{5pt} This is the accepted version of: M. Sul\'ir, M. Ba\v{c}\'ikov\'a, S. Chodarev, J. Porub\"an. Visual augmentation of source code editors: A systematic mapping study. Journal of Visual Languages \& Computing (JVLC), Vol. 49, 2018, pp. 46--59, Elsevier. \url{https://doi.org/10.1016/j.jvlc.2018.10.001}

}
\end{textblock}

\begin{frontmatter}

\title{Visual augmentation of source code editors:\\A systematic mapping study}

\author{Mat\'u\v{s} Sul\'ir\corref{corresponding}}
\ead{matus.sulir@tuke.sk}
\author{Michaela Ba\v{c}\'ikov\'a\corref{}}
\ead{michaela.bacikova@tuke.sk}
\author{Sergej Chodarev\corref{}}
\ead{sergej.chodarev@tuke.sk}
\author{Jaroslav Porub\"an\corref{}}
\ead{jaroslav.poruban@tuke.sk}
\address{Technical University of Ko\v{s}ice, Letn\'a 9, 042 00 Ko\v{s}ice, Slovakia}
\cortext[corresponding]{Corresponding author}

\begin{abstract}
Source code written in textual programming languages is typically edited in integrated development environments or specialized code editors. These tools often display various visual items, such as icons, color highlights or more advanced graphical overlays directly in the main editable source code view. We call such visualizations source code editor augmentation.

In this paper, we present a first systematic mapping study of source code editor augmentation tools and approaches. We manually reviewed the metadata of 5,553 articles published during the last twenty years in two phases -- keyword search and references search. The result is a list of 103 relevant articles and a taxonomy of source code editor augmentation tools with seven dimensions, which we used to categorize the resulting list of the surveyed articles.

We also provide the definition of the term source code editor augmentation, along with a brief overview of historical development and augmentations available in current industrial IDEs.
\end{abstract}

\begin{keyword}
source code editor augmentation\sep integrated development environment (IDE)\sep in situ visualization\sep systematic review\sep survey
\end{keyword}

\end{frontmatter}

\section{Introduction}

Despite decades of research effort, visual programming languages have not obtained a widespread industrial adoption. The majority of programmers write their programs using traditional, textual programming languages \cite{Asenov14envision}. They use a standalone text editor or an integrated development environment (IDE) whose most important component is a textual source code editor. However, upon a closer look at these ``plain text'' editors, we can discover many visual features: from simple syntax highlighting, through underlining of the code violating code conventions, to information about last version control commits in the left margin. We call such visualizations \textit{source code editor augmentation}.

The main purpose of visual source code editor augmentation is to enable spatial immediacy \cite{Ungar97debugging}, i.e. to lower the on-screen distance between the related objects or events. For example, instead of reading the line number of the syntax error and manually navigating to it, the IDE notifies us about its occurrence by underlining the erroneous code directly in the editor.

There exist surveys about software visualization in general \cite{Gracanin05software} and about its various subfields, such as architecture \cite{Shahin14systematic}, algorithm \cite{Simonak13algorithm} or awareness visualization \cite{Storey05use}. Maletic et al. \cite{Maletic02task} presented a task-oriented taxonomy of software visualization. Sutherland et al. \cite{Sutherland16freeform} review ink annotations of digital documents, including program code. Recent systematic mapping studies of other fields related to computer languages include domain-specific languages~\cite{Kosar16dsl} and template-based code generation~\cite{Syriani18systematic}. In our previous work, we surveyed the assignment of metadata to different parts of source code \cite{Sulir17labeling} and included a discussion about the presentation of source code annotations (inside or outside the code). Nevertheless, according to our knowledge, there is no survey available specifically about visual augmentation of source code editors. In this paper, we present a \textit{first systematic mapping study of source code editor augmentation approaches and tools}, analyzing research papers published during the last twenty years.

In section \ref{s:augmentation}, we will define the term source code editor augmentation, provide a brief overview of historical development in this area and mention several augmentations available in current industrial IDEs. We will describe the method used for the systematic mapping study in section \ref{s:method}. The result of our survey is a taxonomy with seven dimensions (section \ref{s:taxonomy}) and a categorized list of the surveyed articles (section \ref{s:approaches}).

\section{Source code editor augmentation} \label{s:augmentation}

In the research area of virtual reality, an augmented reality system ``supplements the real world with virtual objects'' \cite{Azuma01recent}. We would like to apply this terminology to the area of source code editing. In our case, the ``real world'' is represented by the textual source code stored in a file and displayed in a text editor as-is. The ``virtual objects'' are various line decorations, icons, coloring, images, additional textual labels and other visual overlays.

Therefore, we define the term \textit{visual source code augmentation} as an approach which displays additional graphical or textual information directly in plain-text source code. If the code is editable, we can call this also \textit{source code editor augmentation}.

Let us make the definition more precise. First, note that the raw source code displayed in the editor must be the same as the text stored in a file. Projectional editors \cite{Voelter14towards} that use a different editable (visual) and storage representation are, therefore, out of the scope of this article. Furthermore, the augmentation must not remove any part of the displayed source code -- this is analogous to the term ``diminished reality'' in the area of virtual reality\footnote{While some researchers consider diminished reality a subset of augmented reality \cite{Azuma01recent}, we decided to separate these two cases.}.

Second, by the expression ``directly in the code editor'', we mean also the left and right margin of the editor. Note that many IDEs offer many additional views, such as a package explorer or a class navigator. These views are clearly outside the scope of source code editor augmentation.

Alternative terms to ``source code editor augmentation'' are \textit{in situ} software visualization \cite{Harward10in} and source code \textit{annotation} \cite{Swift13visual}. The latter can be easily confused with attribute-oriented programming (e.g., Java annotations \cite{Sulir17exposing}). We decided to use the term \textit{augmentation}, particularly for its correspondence with an existing terminology in the field of computer graphics.

\subsection{Historical view}

Probably the most rudimentary source code editor augmentation feature is \textit{syntax highlighting} (coloring), where each lexical unit is highlighted with a specific color and/or font weight according to its type. One of the first editors supporting real-time source code highlighting was the LEXX editor \cite{Cowlishaw87lexx}, developed in the eighties.

Another useful augmentation feature is \textit{immediate} syntax error \textit{feedback}, pioneered by the Magpie system \cite{Schwartz84incremental}, which incrementally compiled the program being written. Erroneous code fragments were highlighted directly in the editor.

One of the examples from the 90's is ZStep95 \cite{Ungar97debugging}. The expression being currently evaluated was \textit{highlighted} with a border directly in the editor. This augmentation was interactive: a graphical output produced by this expression was displayed in a floating window located next to such a border.

During the last two decades, the advances in computer performance and incremental analysis algorithms enabled the integration of a multitude of augmentation features into mainstream IDEs.

\subsection{Augmentation in industrial IDEs}\label{s:ides}

\begin{figure}
\centering
\includegraphics[scale=0.35]{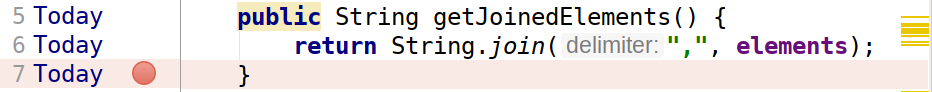}
\caption{A preview of selected augmentation features in an industrial IDE, IntelliJ IDEA.} \label{f:industrial}
\end{figure}

In modern industrial IDEs, visual augmentations are almost omnipresent. For an illustration, Figure~\ref{f:industrial} displays a small selection of them. First, we can see a textual description of the last commit date of each line in the left margin. There is also an icon representing a breakpoint on line 7, which is highlighted also by the red background color. Next, the word ``public'' is highlighted because of a code inspection warning. The parameter name label ``delimiter:'' is only a visual augmentation drawn by the IDE in addition to the code itself. Finally, in the right margin, there is a mini-map of warnings (yellow) in the whole file.

In this section, we will provide an overview of selected visual augmentation possibilities available in current industrial IDEs. We chose the following representatives for the analysis: IntelliJ IDEA 2017.3.5, JetBrains WebStorm 181.4203.9, Visual Studio 2017, Eclipse Oxygen.2 Release (4.7.2) and NetBeans 8.2.
All of them except WebStorm are listed in the top 10 used IDEs worldwide\footnote{https://pypl.github.io/IDE.html (accessed in April 2018 )} and each is regularly used by at least one of the authors of this paper.

A common augmentation in industrial IDEs is the \textit{highlighting} of the current places of interest: a currently evaluated language construct, currently edited line, a paired bracket, all occurrences of the currently refactored language construct in code, etc. Usually, text background color highlighting is used for such features. In the currently edited area, \textit{code indentation} is often marked by vertical lines visually connecting the indented regions in code.

Another large category that has become almost as standard as the previous ones are \textit{hints and warnings}. A light-gray font or text strike-through is applied on unused code, grammar typos or deprecated code. Errors and warnings are marked also with a standard interactive icon in the left editor margin. Hints standardly use orange or yellow and warnings use red color, both for text decoration and for the icons. This kind of augmentation was already used in Eclipse 1.0 in 2001. \textit{Language-specific items}, references and dependencies or another kind of meta-information such as bookmarks, overriding, implementing an interface or recursive calls leverage from the same notation mechanism but usually with a specific icon in the left margin.

All occurrences of different kinds of information, including the aforementioned ones, are usually also marked by clickable, thin color markers in the \textit{right editor margin}, so the programmer is able to quickly navigate to  the specific location in a long source code file. Some IDEs also display a document analysis results icon on top of the right margin.

Apart from the details of visual representation, \textit{debugging} augmentation remained pretty much the same historically. One more recent advance is, for example, the displaying of values currently stored in variables directly in code.

Industrial programming today is all about teamwork, which does not come easy without versioning support -- another standardized feature in common IDEs. Changes in the current version are usually marked by icons or color regions in the left editor margin, e.g. \textit{JetBrains} products use line marking of three types: 1.) a green-colored region for added lines, 2.) a blue-colored region for edited lines and 3.) grey triangle icons for deleted code sections. Eclipse uses a similar but visually less notable visualization. Visual Studio, however, uses a different, more recent approach: interactive team activity info markers \textit{on top of code lines}.

Many recent advances in source code editor augmentation emerge mainly in connection with a specific programming language or technology. Considering the priority of important features such as syntax highlighting, they usually include simpler and usually less notable visualizations such as: underlined text for \textit{HTTP links} used in code (Visual Studio); a \textit{run icon} in the left editor margin, usually placed next to a main class, main method, or a runnable script (JetBrains products); color coding of HTML tags background according to their \textit{level of nesting}; or colored rectangles in left margin next to \textit{color codes} used in CSS-based languages. Recently, IntelliJ IDEA introduced smaller annotations such as displaying \textit{method parameter names} in method calls, marked by a light-grey text before a parameter value (Figure~\ref{f:industrial}, line 6).

\section{Method} \label{s:method}

Now we will describe the method used for a systematic mapping study we conducted. A systematic mapping study is a form of a systematic review with more general research questions, aiming to provide an overview of the given research field.

\subsection{Research questions}

We were interested in the following two research questions:

\begin{itemize}
\item \textbf{RQ1:} What source code editor augmentation tools are described in the literature?
\item \textbf{RQ2:} How can they be categorized?
\end{itemize}

\subsection{Selection criteria}

Inclusion and exclusion criteria are an important part of every systematic review. In our study, all included articles must fulfill the following criteria:

\begin{itemize}
\item It is a journal or conference article published between years 1998--2017 (inclusive).
\item It presents a new tool (or significantly extends an existing one) -- e.g., a desktop application, web application or an IDE/editor plugin.
\item The tool visually augments the source code editor.
\item The article contains a clear textual and graphical description (picture) of the augmentation.
\end{itemize}

At the same time, articles meeting any of the following criteria were excluded:

\begin{itemize}
\item Secondary and tertiary studies, items such as keynotes and editorials, articles for which we could not access a full text.
\item Articles describing graphical and semi-graphical languages.
\item Projectional editing, hiding and replacing existing code with other information.
\item Tools displaying only temporary pop-up windows above code (code completion, tooltips).
\item Standard augmentation present in almost all IDEs (syntax highlighting, standard debugging support, syntax error reporting).
\end{itemize}

\subsection{Search strategy}

\begin{figure}
\centering
\includegraphics[width=0.6\textwidth]{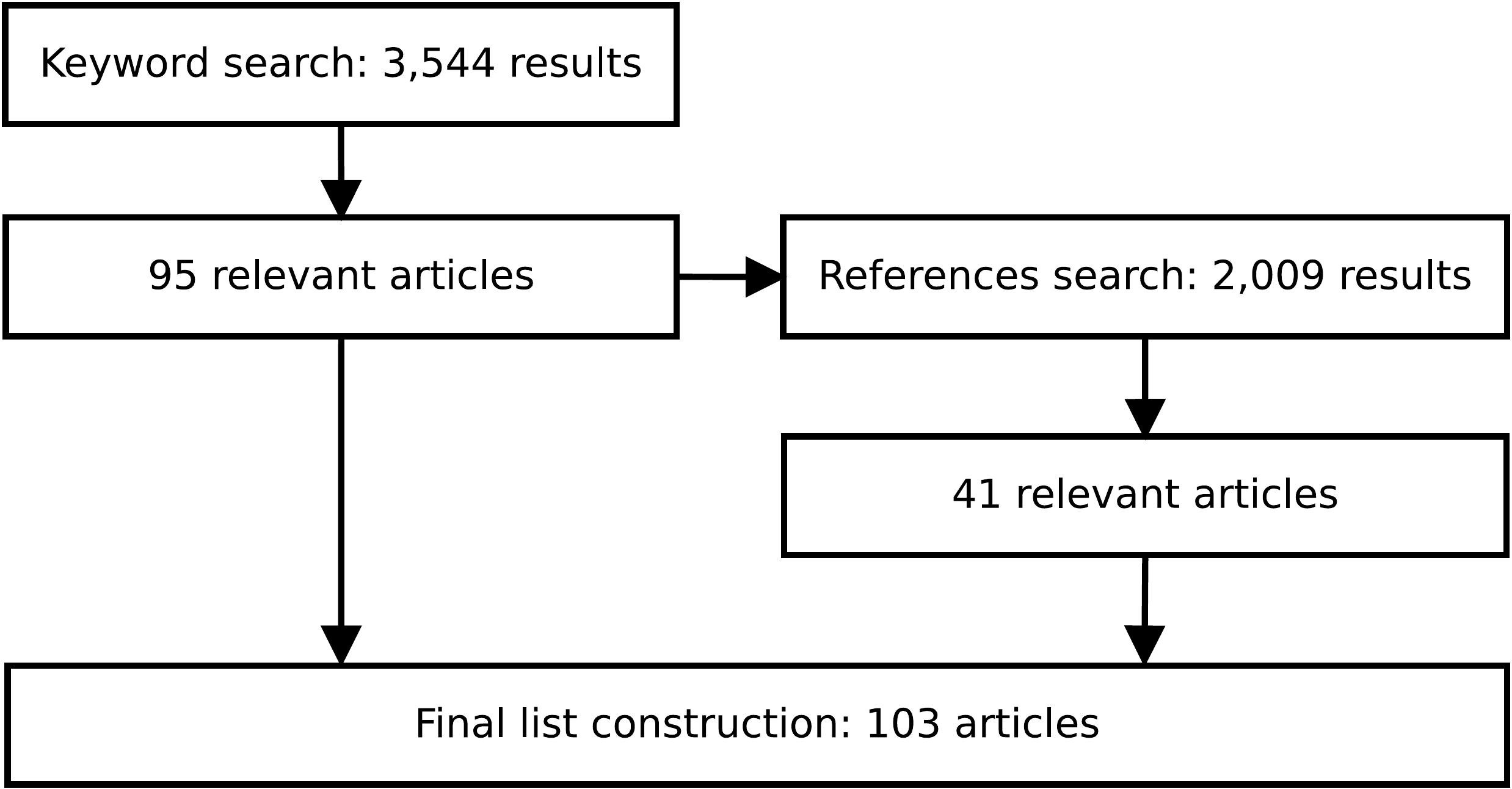}
\caption{A systematic survey search process and article selection overview} \label{f:method}
\end{figure}

To obtain a list of potentially relevant articles, we used a combination of keyword search in digital libraries, forward and backward references search. For an overview of the article search and selection process, see Figure~\ref{f:method}.

\subsubsection{Keyword search}

First, we performed a keyword search in digital libraries to produce a list of potentially relevant articles. The search query was constructed according to the principles suggested by Brereton et al. \cite{Brereton07lessons}: For each element of the study topic (software maintenance, IDE, source code editor, visualization, augmentation), we found synonyms and related words, connected them by the logical ``OR'' operator and then connected the subqueries by the ``AND'' operator. The resulting query is:

\begin{verbatim}
(software OR program) AND (maintenance OR comprehension OR understanding)
AND (tool OR "development environment")
AND ("source code" OR "source file") AND editor
AND (visualize OR visualization OR display)
AND (augment OR augmentation OR annotate OR annotation)
\end{verbatim}

This query was entered into advanced search boxes of four digital libraries: IEEE Xplore\footnote{\url{http://ieeexplore.ieee.org}}, ACM Digital Library (DL)\footnote{\url{http://dl.acm.org}}, ScienceDirect\footnote{\url{http://www.sciencedirect.com}} and Springer Link\footnote{\url{http://link.springer.com}}. These four libraries were chosen because of their popularity in software engineering research \cite{Zhang11identifying}. In the case of ACM DL, we had to accommodate the query to use a different syntax (without changing its semantics). The search was performed on metadata and full texts. We limited the search to the field of computer science, using the possibilities available in individual libraries, e.g., in IEEE Xplore, we excluded articles not contained in the Computer Society DL. We also filtered the entry types to journal/conference articles with a full text available and the publication year to 1998--2017, written in English.

For all found entries we exported metadata such as titles and abstracts. If the given library did not offer the export of abstracts, we downloaded them from Scopus\footnote{\url{http://www.scopus.com}}. After removing duplicate items, this phase produced a list of 3,544 articles.

Then, three of the authors each were given a subset of the articles to evaluate. The researchers manually decided which of these articles fulfill the selection criteria -- first based on the title and abstract review and then, if the article seemed potentially relevant, by skimming the full text. The result was a list of 95 selected articles.

\subsubsection{References search}

Next, we tried to find relevant articles by examining the references of the 95 articles identified in the previous phase (keyword search). We were interested both in backward references (literature cited in these articles) and forward references (newer papers citing these articles). To export the lists of backward and forward references along with all necessary metadata, we used Scopus. Similarly to the previous phase, we limited the search to English computer science journal/conference articles published between 1998 and 2017, whenever these options were available.

Two of the articles were not indexed in Scopus and for these, we extracted the backward references manually from the full texts; forward references were exported from the digital libraries of the articles' origin (ACM DL and ScienceDirect).

We merged duplicate entries between forward and backward reference lists. Next, we removed papers already present in the keyword search list. This resulted in 2,009 unique articles.

This set of articles was again manually evaluated by the authors, similarly to the previous phase -- first according to the metadata and then a subset of them considering the full texts. The result was a list of 41 more papers. The search was non-recursive (i.e., not snowballing) since we did not extract the reference lists of these new articles again.

\subsubsection{Final list construction}

By combining the results from keyword search and references extraction, we obtained 136 relevant articles. However, 31 of them were describing the same tools as the other ones, only in a different research stage (idea papers, evaluations, etc.). Two articles were classified as ``not relevant'' after a more careful inspection.

Therefore, the final set consists of 103 relevant articles.

\subsection{Data extraction}

For each article, we tried to find supplementary material on the web, such as manuals, screenshots, videos and executable tool downloads. We also extracted a tool name from every article; if the tool was unnamed, we devised a suitable pseudonym based on the keywords often used in the paper. In rare cases when one article presented more tools, we selected the most relevant one.

Three of the authors were each given a portion of the tools to classify. Using the full texts of the articles and available supplementary information, they tried to find similar and distinguishing characteristics of the tools. In this first round, each researcher gradually formed his/her own taxonomy (which could be freely inspired by others). A taxonomy consisted of a number of dimensions (e.g., ``visualization type'') with their attributes (e.g., ``icon'', ``text''). One tool pertained to one or more attributes of every dimension.

Next, two of the classifying authors and the fourth author merged the taxonomies on a personal meeting. The three researchers then re-classified their tools accordingly. Finally, after a discussion, we renamed, split and merged some of the attributes.

The final result is: 1.) a taxonomy with 7 dimensions, each consisting of 2--6 attributes, and 2.) a categorization of all analyzed tools according to this taxonomy.

\subsection{Threats to validity}

Now we will discuss the validity threats of individual systematic mapping study parts.

Since the article selection process was performed by three authors, the inclusion and exclusion criteria could be comprehended differently by individual researchers. However, we intensively discussed the criteria in the team, along with the examples of individual tools matching and not matching the criteria to lower the variability to a minimum.

The terminology in the area (e.g., the expression ``source code editor augmentation'') is not yet standardized, so we could miss multiple relevant articles. An example of such an article is the vsInk tool paper \cite{Sutherland13vsink}, describing ink annotations for editable source code.

Although the keyword search results were not compared to any quasi-gold standard, we constructed it also by considering the results of our previous survey \cite{Sulir17labeling} that included multiple articles present in the current review.

During the backward references search in Scopus, we did not include so-called secondary documents, i.e. articles not indexed by Scopus but found in the reference lists. The reason is that this option resulted in many entries with incomplete bibliographical information, which would complicate the selection process.

Both the article selection and data extraction were performed by three researchers, each evaluating a non-overlapping portion of articles. Nevertheless, questionable items were discussed until a consensus was reached. Furthermore, after each part, one of the authors checked the list of all relevant articles to ensure the overall quality and to exclude inappropriate items.

Tools are often a neglected aspect of research, so some papers might not describe all available features of the given tool. Note that we do not draw any precise quantitative conclusions about the analyzed articles, such as exact percentages of tools in each category. The main purpose of this paper is to offer an overview of the field, where each of the analyzed tools acts as an example of particular augmentations and a reference for further information.

\section{Taxonomy} \label{s:taxonomy}

To answer \textbf{RQ2}, we present the resulting taxonomy of source code editor augmentation in Table~\ref{t:taxonomy}. In this section, we will describe the individual dimensions and their attributes, while mentioning representative examples of tools.

\begin{table}
\vspace*{-2cm}
\caption{The taxonomy of source code editor augmentation}
\label{t:taxonomy}
\hspace*{-1cm}
\centering
\small
\renewcommand{\arraystretch}{1.3}
\begin{tabular}{|l|p{13cm}|} \hline
\textbf{Dimension} & \textbf{Description} \\
\textit{and its attributes} & ~ \\ \hline

\textbf{Source} & Where does the data representing the augmentation come from? \\
\textit{code} & Results of static source code analysis. \\
\textit{runtime} & Results of the program execution; dynamic program analysis. \\
\textit{human} & Manually entered information, previously present only in the human mind. \\
\textit{interaction} & Interaction patterns of a single developer in the IDE or a similar tool. \\
\textit{collaboration} & Behavior of multiple developers/users and their artifacts (like VCS commits). \\ \hline
\textbf{Type} & Data of what type does the augmentation directly represent? \\
\textit{boolean} & The augmentation can be only present or non-present. \\
\textit{fixed enumeration} & One of a set of possible categorical values, the categories are pre-defined by the tool itself. \\
\textit{variable enumeration} & One of a set of possible categorical values, the number of categories is not fixed. \\
\textit{number} & Numeric values (numbers, time, etc.). \\
\textit{string} & A text string. \\
\textit{object} & Another data type (an image, a complex structure, etc.). \\ \hline
\textbf{Visualization} & What does the augmentation look like? \\
\textit{color} & A simple area filled with a background color, a color bar or text foreground color. \\
\textit{decoration} & Decoration of the code, such as underline, overline or border. \\
\textit{icon} & A small rectangular graphical object. \\
\textit{graphics} & More complicated graphical representations -- charts, arrows, diagrams, photos, etc. \\
\textit{text} & An inserted text string (including a number written as a text). \\ \hline
\textbf{Location} & Where is the visual augmentation displayed? \\
\textit{left} & To the left of the source code - usually in the left margin. \\
\textit{in code} & Directly in the code. \\
\textit{right} & It is right-aligned. \\ \hline
\textbf{Target} & To what is the augmentation visually assigned? \\
\textit{line} & One line. \\
\textit{line range} & Multiple lines, but without distinguishing characters in individual lines. \\
\textit{character range} & A number of characters (on one line or multiple lines). \\
\textit{file} & The whole file. \\ \hline
\textbf{Interaction} & How can we interact with the augmentation? \\
\textit{popover} & A tooltip or a popup with additional information can appear at the place where the augmentation is displayed. \\
\textit{navigate} & Performing an action directly on the augmentation navigates us to another code location, window, website, etc. \\
\textit{change} & We can edit the displayed information directly by manipulating the augmentation (drag\&drop, write, expand, collapse, highlight). \\
\textit{none/unknown} & The augmentation does not offer interaction possibilities or they were not mentioned by the authors. \\ \hline
\textbf{IDE} & For what IDE/editor is the augmentation implemented? \\
\textit{existing} & An existing IDE/editor is extended. \\
\textit{custom} & A tool created by the authors specifically for the purposes described in the article (or prior works). \\ \hline

\end{tabular}
\end{table}

\subsection{Source}

The \textbf{source} dimension denotes where the data representing the augmentation were originally available before they were visually assigned to a part of source code. This dimension is similar to the one presented in our previous work \cite{Sulir17labeling}.

Tools categorized as \textit{code} analyze the source code of a system without executing it, i.e., using static analysis. This can range from simple markers about the presence of an MPI (Message Passing Interface) function call on the given line \cite{Watson06developing} to a sophisticated calculation of potential deployment costs using static analysis \cite{Leitner16modelling}. A very common topic is clone detection (e.g., CloneTracker \cite{Duala-Ekoko07tracking}, SimEclipse \cite{Uddin15towards}, SourcererCC-I \cite{Saini16sourcerercc}) and warnings about bad smells (JDeodorant \cite{Tsantalis08jdeodorant}, Stench Blossom \cite{Murphy-Hill10interactive}).

Data useful for augmentation can be collected by an execution of the program, using some form of dynamic analysis. These tools are marked as \textit{runtime}.

The majority of \textit{runtime}-sourced tools collect the data during an execution and display them afterward, when the program has stopped. For instance, IDE sparklines \cite{Beck13visual} are small graphics displaying the progress of numeric variable values over time. A common application is performance profiling (e.g., In situ profiler \cite{Beck13in}). An important problem of this kind of tools is data invalidation: if the source code is modified after the data were collected, the runtime information may not be valid anymore and the program must be run again. This issue is rarely discussed in the reviewed articles. A notable exception is Senseo \cite{Roethlisberger12exploiting}, where the authors explicitly state the data pertaining to the modified method and its dependencies are always invalidated.

Another type of tools utilizing the \textit{runtime} source, live programming environments such as Impromptu HUD \cite{Swift13visual} and Gibber \cite{Roberts14gibber}, display the augmentation in real time as the program is executing.

For tools utilizing the \textit{human} source, a programmer must purposefully enter the data with a sole intention that they will be used by the augmentation. Typical examples are social bookmarks in Pollicino \cite{Guzzi11collective}, tags in TagSEA \cite{Storey09how} or manual concern-to-code mappings in Spotlight \cite{Revelle05understanding}.

On the other hand, \textit{interaction} data are collected automatically, possibly without the developer even knowing it (although that would be unethical). A typical example is local code change tracking for the purpose of clone detection (CnP \cite{Jablonski10aiding}, CSeR \cite{Jacob10actively}).
Some tools also collect data from external applications, for example, HyperSource \cite{Hartmann11hypersource} tracks the user's web browsing activity while editing a code part and then augments that part of the code with potentially relevant browsing history.

As software engineering is not an individual activity, collaboration data are formed naturally as the team communicates and cooperates. These data are utilized by \textit{collaboration} tools. Some of these tools offer an analysis of existing artifacts. For instance, Rationalizer \cite{Bradley11supporting} displays last VCS (version control system) commit times, authors, commit messages and issues related to each source code. The second common category is represented by real-time collaborative source code editors, such as Saros \cite{Salinger10saros}, Collabode \cite{Goldman11real} and IDEOL \cite{Dang14educo}.

Many tools use a combination of multiple methods. For example, Historef \cite{Hayashi15historef} automatically collects \textit{interaction} data -- local source code edits -- in the background and then allows the developer to manually merge, split or reorder these edits (the \textit{human} source) while highlighting the changes with different colors.

\subsection{Type}

The type of data represented directly by the visual augmentation is denoted by the \textbf{type} dimension. This can be regarded as an extension of the classical information visualization theory \cite{Card97structure}, which recognizes nominal, ordered and quantitative data.

The \textit{boolean} type means that the augmentation is either present or not, without distinguishing any visual variations. Two most common boolean augmentations are one-type marker icons and one-color code highlights. For example, Remail \cite{Bacchelli11miler} displays a marker icon in the left margin on each line which has a related e-mail available. The iXj plugin \cite{Boshernitsan07aligning} highlights all code matching the given transformation pattern with a green background. Note that the boolean type is often insufficient to display all necessary information, so it is commonly combined with interaction possibilities (e.g., tooltips) or displayed only for a limited amount of time -- after a specific action.

Some tools use a list of predefined values for displaying different augmentation notations. They can either be firmly stored as a finite list in the tool's data bank, i.e. it's a \textit{fixed enumeration}; or the number of categories can be changed any time by the user or by the tool itself, i.e. \textit{variable enumeration}.

An example of \textit{fixed enumeration} augmentation can be seen in Syde \cite{Hattori10syde} where a red left-margin highlight means a severe collaboration conflict and the yellow color represents moderate conflicts.

A very common application of \textit{variable enumeration} augmentation is an assignment of a different color for every concern or feature in the source code. This technique is used with variations in at least seven tools: Spotlight \cite{Revelle05understanding}, CIDE \cite{Kaestner08granularity}, ArchEvol \cite{Nistor09explicit}, IVCon \cite{Saigal09inline}, FeatureIDE \cite{Meinicke16featureide}, FLOrIDA \cite{Andam17florida} and xLineMapper \cite{Zheng17mapping}. Another common approach is to assign a specific color for each developer working on a piece of code, used e.g., in ATCoPE \cite{Fan12atcope}. However, there is a problem with such assignments: The color has only a limited number of easily distinguishable levels \cite{Moody09physics}. Without additional visualizations and interactions, the utility of such approaches would be questionable if the number of elements in the enumeration (features, people, etc.) raised beyond a reasonable limit.

If the augmentation is represented by a numeric value, e.g. a number of function calls or time, then we use the type \textit{number}. For example, Clepsydra \cite{Harmon07interactive} calculates and displays worst-case execution times for individual code segments and Theseus \cite{Lieber14addressing} displays the number of function calls in a JavaScript file while the application is running in a browser.

The \textit{string} type represents a text string which is not enumerated, i.e., we cannot name all its possible values. For instance, the Live coding IDE \cite{Kramer14how} shows live values of variables in the source code editor.

Any other displayed information, such as an image or a complex data structure belongs to the \textit{object} type. One such example is SE-Editor \cite{Schugerl09beyond}, which embeds web pages and images directly into the editor.

\subsection{Visualization}

\begin{figure}
\centering

\begin{subfigure}[b]{0.48\textwidth}
\includegraphics[width=\textwidth]{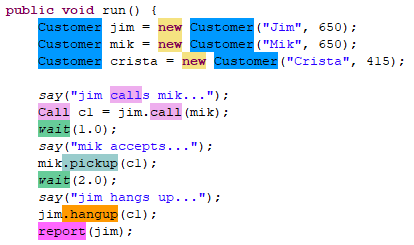}
\caption{\textit{color}} \label{f:color}
\end{subfigure}
~
\begin{subfigure}[b]{0.48\textwidth}
\includegraphics[width=\textwidth]{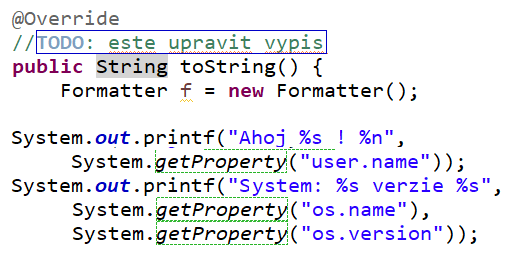}
\caption{\textit{decoration}} \label{f:decoration}
\end{subfigure}
\par\bigskip
\begin{subfigure}[b]{0.48\textwidth}
\includegraphics[width=\textwidth]{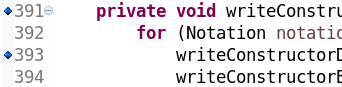}
\caption{\textit{icons}} \label{f:icon}
\end{subfigure}
~
\begin{subfigure}[b]{0.48\textwidth}
\includegraphics[width=\textwidth]{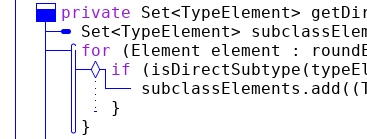}
\caption{\textit{graphics}} \label{f:graphics}
\end{subfigure}
\par\bigskip
\begin{subfigure}[b]{0.50\textwidth}
\includegraphics[width=\textwidth]{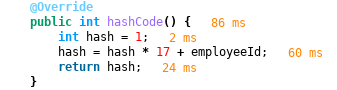}
\caption{\textit{text}} \label{f:text}
\end{subfigure}

\caption{Tools with various visualization kinds:
(a) Aspect Browser \cite{Griswold01exploiting} uses text background \textit{colors} to identify different aspects of code according to user-defined patterns;
(b) eMoose \cite{Dekel09improving} displays a solid border around TODO comments and adds dashed border \textit{decorations} to code representing associated usage directives or reverse TODO references;
(c) Pollicino \cite{Guzzi11collective} uses \textit{icons} to represent bookmarks; 
(d) jGRASP \cite{CrossII11combining} displays control structure diagram \textit{graphics} directly in code; and 
(e) Clepsydra \cite{Harmon07interactive} calculates worst-case execution times and shows them as \textit{textual labels} next to individual code lines. } \label{f:visualization}
\end{figure}

The \textbf{visualization} dimension relates to the visual appearance of the augmentation. For an illustration of various visualization kinds, see Figure~\ref{f:visualization}.

Color is a very perceivable visual sign and at the same time the most simple one, applicable to both left or right editor bar and the code text itself. Color bars in the margins, text background and foreground color in the editor belong to the \textit{color} category. A tool can utilize one color (Traces view \cite{Alsallakh12visual}), multiple distinct colors (Jigsaw \cite{Cottrell08semi}) or a color spectrum. When utilizing the color spectrum, tools can change the hue (vLens \cite{Li13calculating}), saturation (CodeMend \cite{Rong16codemend}) or brightness (CnP \cite{Jablonski10aiding}) according to a numeric value. Note that the code foreground color is often reserved for syntax highlighting. Therefore its use is very limited -- we encountered only one such tool, IVCon \cite{Saigal09inline}, which does not use syntax highlighting at all.

A less notable highlight is usually represented by a text underline, overline, or border, belonging to the \textit{decoration} category. A dashed border surrounding a code part is used in multiple tools, for example, in eMoose \cite{Dekel09improving} it represents code with associated usage directives, such as threading limitations or reverse TODO references. The XSS marker plugin \cite{Bathia11assisting} utilizes a standard Eclipse visualization form, red squiggle underline, but with a different meaning -- instead of denoting compilation errors, it warns about cross-site scripting vulnerabilities.

\textit{Icons} are usually used in editor left bars as small, mostly rectangular and simple graphics. They can visually represent the metaphor of the given tool, e.g., in CodeBasket \cite{Biegel15codebasket}, an egg icon in the left margin represents a ``code basket egg'', i.e. a part of the programmers mental model. In other cases they are only attention-catching symbols, e.g., in DSketch \cite{Cossette10dsketch}, red square icons represent lines containing matches of dependency analysis. Occasionally, icons are located directly in the code editor (instead of the left margin) as in the ALVIS Live! tool \cite{Hundhausen07what}.

More complicated graphical representations such as graphs, charts, arrows, diagrams, photos, etc. belong to the \textit{graphics} group. For example, FluidEdt \cite{Ou15interactive} displays heap graphs in the left margin. I3 \cite{Beck15rethinking} offers search similarity and change history views in form of small charts directly in the editor. The Fractional ownership tool \cite{Mueller15in} displays code authorship history as multi-colored stripes for each code line in the right editor part. A prototype ``Error notifications'' IDE \cite{Barik14compiler} shows visual descriptions of compiler errors -- e.g., in case of a name clash, two relevant code elements are visually connected with a line. DrScheme \cite{Findler02drscheme} connects individual uses of a selected language construct in the code by arrows. Finally, the Cares plugin \cite{Guzzi12facilitating} displays photos of code-related people in the editor.

Any piece text added to the editor is considered to be a \textit{text} augmentation. It can be subtle, as in CeDAR \cite{Tairas12increasing}, where a clone region is labeled with a number in the corner: e.g., ``Clone 1''. Another example is the ``ghost comments'' technique of Moonstone \cite{Kistner17moonstone} non-editable comment-like labels at the end of each line, or the Clepsydra tool's \cite{Harmon07interactive} line labeling with execution times. On the other hand, Code portals \cite{Breckel16embedding} embed significant portions of complementary texts into the code editor, such as relevant source code parts or visualizations of line differences.

\subsection{Location}

\begin{figure}
\centering

\begin{subfigure}[b]{0.38\textwidth}
\includegraphics[width=\textwidth]{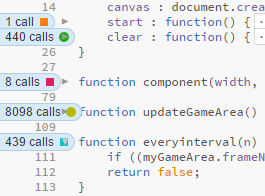}
\caption{\textit{left}} \label{f:left}
\end{subfigure}
~
\begin{subfigure}[b]{0.40\textwidth}
\includegraphics[width=\textwidth]{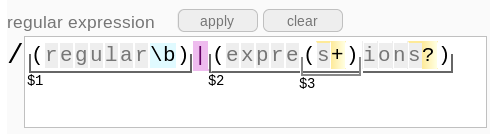}
\caption{\textit{in code}} \label{f:in-code}
\end{subfigure}
\par\bigskip
\begin{subfigure}[b]{0.75\textwidth}
\includegraphics[width=\textwidth]{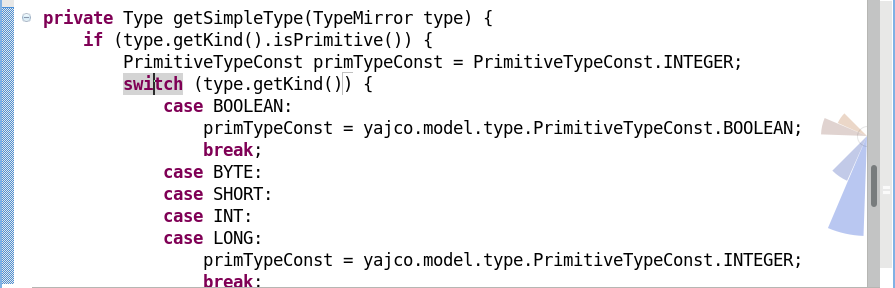}
\caption{\textit{right}} \label{f:right}
\end{subfigure}

\caption{Tools with various location kinds:
(a) Theseus \cite{Lieber14addressing} displays the number of function calls in a JavaScript code in the \textit{left} editor part;
(b) RegViz \cite{Beck14regviz} highlights regular expressions structure directly \textit{in code}; and
(c) Stench Blossom's \cite{Murphy-Hill10interactive} ambient smell detector that lives on the \textit{right} edge of the program editor.}\label{f:location}
\end{figure}

The \textbf{location} dimension denotes the position of the visual augmentation in the editor, which can either be to the \textit{left} of the source code, displayed directly \textit{in code}, or placed in the \textit{right} editor part, aligned to the right editor margin. See Figure~\ref{f:location} for an illustration.

The \textit{left} margin, alternatively called gutter or ruler, traditionally displays line numbers and simple icons (e.g., iMaus \cite{Kawrykow09improving}). However, it can show also thin color bars (Diver \cite{Myers10using}), thicker color fills (HyperSource \cite{Hartmann11hypersource}), ``pills'' containing text (Theseus \cite{Lieber14addressing}) or even a list of callers (Stacksplorer \cite{Karrer11stacksplorer}).

Among tools displaying augmentation \textit{in code}, we can distinguish multiple cases. The jGRASP tool \cite{CrossII11combining} displays a control structure diagram in the indentation area of source code containing only tabs or spaces so it does not visually overlap with the text. DrScheme \cite{Findler02drscheme} displays arrows directly above code, covering small parts of it; however, this augmentation is only temporary displayed on mouse hover, which makes it acceptable. ARCC \cite{Nunez17arcc} draws augmentation behind the code in the form of a background color. SketchLink \cite{Baltes14linking} displays icons directly in the code, but only at the end of lines. Finally, RegViz \cite{Beck14regviz} slightly reflows the text so that the visual annotations do not overlap it.

\textit{Right} augmentation components are pinned to the editor edges, which means that if the IDE window resizes (changes its width), the components move along with its edges. ``Color chips'' in CoderChrome \cite{Harward10in}, displayed in the right part of the editor, are an example of such a visualization. Their meaning can be configured, e.g., to mark starts and ends of code blocks. Because the right margin is usually not the central programmer's focus point, it is an ideal place to implement ambient, non-disturbing augmentations -- such as the Stench Blossom \cite{Murphy-Hill10interactive} code smell visualization. Since this location is near the scrollbar, it is also useful for general overviews: heatmap-colored icons in Senseo \cite{Roethlisberger12exploiting} referring to available runtime information in the whole file.

\subsection{Target}

\textbf{Target} denotes the code construct to which the augmentation is visually assigned.

\textit{Line} augmentations are bound to a single line of code, \textit{line range} to multiple consecutive lines. Augmentations displayed in the left margin are practically always assigned to a \textit{line} or a \textit{line range}. For instance, each left-margin crash analysis icon in CSIclipse \cite{Ohmann15csiclipse} is related to a particular \textit{line} (e.g., ``line may have been executed''). \textit{Line range} augmentations are typically denoted as color bars in the left margin (Featureous \cite{Olszak12remodularizing}) or code background color spanning whole lines, up to the right margin (EnergyDebugger \cite{Banerjee16debugging}).

On the other hand, a \textit{character range} augmentation can mark an arbitrary selection of individual characters. The augmentation can either stay on one line (e.g., a border decoration in ChangeCommander \cite{Fluri08recommending}) or potentially span multiple lines (background color highlighting in CodeGraffiti  \cite{Lichtschlag14codegraffiti}).

\textit{Line} and \textit{character range} augmentations are often used in tandem: A left margin icon marks the line of interest while the decoration (TexMo \cite{Pfeiffer12texmo}) or background color (Gilligan \cite{Holmes07task}) highlights a more specific source code range in the given line.

If the augmentation relates to the whole code in the currently opened file, then it belongs to the \textit{file} category. Such a visualization can spatially correspond to source code parts (jGRASP \cite{CrossII11combining}), or its placement can be solely a matter of style (Gibber \cite{Roberts14gibber}).

\subsection{Interaction}

An icon, text or any other graphics or representation is usually not sufficient to give the user enough information about the augmentation. Tools usually provide some way of displaying more data about the particular case, which is usually initiated by the user interacting with a graphical component representing the augmentation. The \textbf{interaction} dimension describes the ways of how it is possible to interact with the displayed augmentation.

\begin{figure}
\centering
\includegraphics[width=0.5\textwidth]{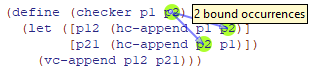}
\caption{DrScheme \cite{Findler02drscheme}, currently named DrRacket -- a tool utilizing the \textit{popover} interaction in form of a tooltip denoting the number of bound occurrences of particular constructs in code.} \label{f:popover}
\end{figure}

The most common interaction (if there is any) is a case when a \textit{popover} (a tooltip or a popup) with additional information displays after a mouse click or a mouse cursor hovering over the augmentation. This can be a simple textual tooltip, as depicted in Figure~\ref{f:popover} (DrScheme \cite{Findler02drscheme}), or a multi-line formatted text with a picture and multiple clickable actions (Cares \cite{Guzzi12facilitating}).

If there is a lot of additional data to display for the augmentation and they would not fit into a popup, clicking or double-clicking on the augmentation component usually fires a \textit{navigation} event, which selects an item in another IDE view/window (traceability links in Morpheus \cite{Eyl17traceability}) or displays the information in some external program, e.g. a web browser (a bug report page in the case of Rationalizer \cite{Bradley11supporting}). If the augmentation is somehow related to other parts of the code, the user is navigated directly to those code parts in the same or another file -- e.g., control-flow and data-flow hyperlinks in Flower \cite{Smith17flower}.

Some tools enable direct manipulation of the augmentation itself using mouse or keyboard interaction, which \textit{changes} the displayed information or its appearance. Code portals \cite{Breckel16embedding} allow the developer to edit the displayed inline texts. The FixBugs tool \cite{Barik16from} offers drag\&drop refactoring capabilities directly in the source code editor. In Fluid views \cite{Desmond06fluid}, the interaction happens in two steps: First, an unobtrusive visual cue changes to an underline decoration on a mouse hover. After clicking it, the augmentation fully expands and the related code is displayed inside the editor.

If the augmentation component is not interactive in any way or the authors of the tool did not mention any possibilities of interaction in their paper or any supplementary materials, it is classified as \textit{none/unknown}.

\subsection{IDE}

The last dimension denotes the \textbf{IDE} or editor, for which the augmentation is implemented. The solutions are either implemented as plugins into an \textit{existing} IDE or code editor, or they are completely new \textit{custom} environments created by the authors specifically for the purposes described in their article.

Most of the tools we identified were implemented as Eclipse plugins, which is obvious given the popularity and open architecture of this IDE.
Some of them were created for other more or less popular IDEs such as Visual Studio (GhostFactor \cite{Ge14manual}), IntelliJ IDEA (wIDE \cite{Murolo17improved}) or Brackets IDE (Theseus \cite{Lieber14addressing}). Examples of custom tools created from scratch are Omnicode \cite{Kang17omnicode}, Shared-code editor \cite{Lee17exploring}, Code portals \cite{Breckel16embedding} and Code Bubbles\cite{Bragdon10code}.

\section{Tools} \label{s:approaches}

As an answer to \textbf{RQ1}, we provide a complete list of the surveyed tools in Tables~\ref{t:tools1} and \ref{t:tools2}. The rows are individual tools, sorted alphabetically by the tool name. Each column represents an attribute of the corresponding dimension.

\begin{table}
\vspace*{-3cm}
\caption{Augmentation tools}
\label{t:tools1}
\hspace*{-2cm}
\setlength{\tabcolsep}{0.38em}
\renewcommand{\arraystretch}{1.14}
\centering
\footnotesize
\begin{tabular}{|l|l|ccccc|cccccc|ccccc|ccc|cccc|cccc|c|} \hline
\multirow{2}{*}{\textbf{Tool}}
 & \multirow{2}{*}{\textbf{Year}}
 & \multicolumn{5}{c|}{\textbf{Source}}
 & \multicolumn{6}{c|}{\textbf{Type}}
 & \multicolumn{5}{c|}{\textbf{Visualization}}
 & \multicolumn{3}{c|}{\textbf{Location}}
 & \multicolumn{4}{c|}{\textbf{Target}}
 & \multicolumn{4}{c|}{\textbf{Interaction}}
 & \multicolumn{1}{c|}{\textbf{IDE}} \\ \cline{3-30}

~ & ~ & \h{code} & \h{runtime} & \h{human} & \h{interaction} & \h{collaboration} & \h{boolean} & \h{fixed enum.} & \h{variable enum.} & \h{number} & \h{string} & \h{object} & \h{color} & \h{decoration} & \h{icon} & \h{graphics} & \h{text} & \h{left} & \h{in code} & \h{right} & \h{line} & \h{line range} & \h{character range} & \h{file} & \h{popover} & \h{navigate} & \h{change} & \h{none/unknown} & \h{existing} \\ \hline

ALVIS Live! \cite{Hundhausen07what} & 2007 & \y & \n & \n & \n & \n & \y & \n & \n & \n & \n & \n & \n & \n & \y & \n & \n & \n & \y & \n & \y & \n & \n & \n & \n & \y & \n & \n & \n \\
ARCC \cite{Nunez17arcc} & 2017 & \y & \n & \n & \n & \n & \y & \n & \y & \n & \n & \n & \y & \y & \n & \n & \n & \y & \y & \n & \y & \n & \y & \n & \y & \n & \n & \n & \y \\
ArchEvol \cite{Nistor09explicit} & 2009 & \n & \n & \y & \n & \n & \n & \n & \y & \n & \n & \n & \y & \y & \n & \n & \n & \y & \y & \n & \n & \y & \n & \n & \y & \n & \n & \n & \y \\
ASIDE \cite{Xie11aside} & 2011 & \y & \n & \y & \n & \n & \y & \y & \n & \n & \n & \n & \y & \y & \y & \n & \n & \y & \y & \n & \y & \n & \y & \n & \y & \n & \y & \n & \y \\
Aspect Browser \cite{Griswold01exploiting} & 2001 & \y & \n & \y & \n & \n & \n & \n & \y & \n & \n & \n & \y & \n & \n & \n & \n & \n & \y & \n & \n & \n & \y & \n & \n & \y & \n & \n & \y \\\hline
ATCoPE \cite{Fan12atcope} & 2012 & \n & \n & \n & \n & \y & \n & \n & \y & \n & \n & \n & \y & \n & \n & \n & \n & \y & \y & \n & \n & \y & \n & \n & \n & \n & \n & \y & \y \\
BeneFactor \cite{Ge12reconciling} & 2012 & \y & \n & \n & \y & \n & \y & \n & \n & \n & \n & \n & \n & \n & \y & \n & \n & \y & \n & \n & \y & \n & \n & \n & \y & \n & \y & \n & \y \\
Cares \cite{Guzzi12facilitating} & 2012 & \n & \n & \n & \n & \y & \n & \y & \n & \n & \n & \y & \n & \n & \n & \y & \n & \n & \n & \y & \n & \n & \n & \y & \y & \y & \n & \n & \y \\
CeDAR \cite{Tairas12increasing} & 2012 & \y & \n & \n & \y & \n & \y & \n & \n & \n & \y & \n & \y & \n & \n & \n & \y & \n & \y & \n & \n & \y & \n & \n & \n & \y & \n & \n & \y \\
ChangeCommander \cite{Fluri08recommending} & 2008 & \y & \n & \n & \n & \y & \y & \n & \n & \n & \n & \n & \n & \y & \y & \n & \n & \y & \y & \n & \n & \n & \y & \n & \y & \n & \n & \n & \y \\\hline
Cheetah \cite{Do17cheetah} & 2017 & \y & \n & \n & \n & \n & \n & \y & \n & \n & \n & \n & \y & \n & \y & \n & \n & \y & \y & \n & \y & \n & \n & \n & \y & \n & \n & \n & \y \\
CIDE \cite{Kaestner08granularity} & 2008 & \n & \n & \y & \n & \n & \n & \n & \y & \n & \n & \n & \y & \n & \n & \n & \n & \n & \y & \n & \n & \n & \y & \n & \y & \n & \n & \n & \y \\
Clepsydra \cite{Harmon07interactive} & 2007 & \y & \n & \n & \n & \n & \n & \n & \n & \y & \n & \n & \n & \n & \n & \n & \y & \n & \y & \n & \y & \n & \n & \n & \n & \n & \n & \y & \y \\
CloneTracker \cite{Duala-Ekoko07tracking} & 2007 & \y & \n & \n & \n & \n & \y & \n & \n & \n & \n & \n & \n & \n & \y & \n & \n & \y & \n & \n & \y & \n & \n & \n & \y & \n & \n & \n & \y \\
CnP \cite{Jablonski10aiding} & 2010 & \n & \n & \n & \y & \n & \y & \n & \y & \y & \n & \n & \y & \n & \y & \n & \n & \y & \n & \n & \y & \y & \n & \n & \y & \n & \n & \n & \y \\\hline
Code Bubbles \cite{Bragdon10code} & 2010 & \n & \n & \n & \y & \n & \y & \n & \n & \n & \n & \n & \y & \n & \n & \y & \n & \y & \y & \y & \y & \n & \n & \n & \n & \n & \y & \n & \n \\
Code Orb \cite{Lopez11code} & 2011 & \n & \n & \n & \n & \y & \n & \n & \n & \y & \n & \n & \y & \n & \n & \n & \n & \y & \n & \n & \y & \n & \n & \n & \n & \y & \n & \n & \y \\
Code portals \cite{Breckel16embedding} & 2016 & \y & \y & \n & \n & \y & \n & \n & \y & \n & \y & \n & \y & \y & \n & \n & \y & \n & \y & \n & \y & \n & \y & \n & \n & \n & \y & \n & \n \\
CodeBasket \cite{Biegel15codebasket} & 2015 & \n & \n & \y & \n & \n & \y & \n & \n & \n & \n & \n & \n & \n & \y & \n & \n & \y & \n & \n & \y & \n & \n & \n & \n & \y & \n & \n & \y \\
CodeGraffiti \cite{Lichtschlag14codegraffiti} & 2014 & \n & \n & \y & \n & \n & \n & \n & \y & \n & \n & \n & \y & \n & \n & \n & \n & \y & \y & \n & \y & \n & \y & \n & \n & \y & \n & \n & \y \\\hline
Codelink \cite{Toomim04managing} & 2004 & \y & \n & \y & \y & \n & \y & \y & \n & \n & \n & \n & \y & \n & \n & \n & \n & \n & \y & \n & \n & \n & \y & \n & \n & \n & \y & \n & \y \\
CodeMend \cite{Rong16codemend} & 2016 & \y & \n & \n & \n & \y & \n & \y & \n & \n & \n & \n & \y & \n & \n & \n & \n & \n & \y & \n & \y & \n & \y & \n & \n & \y & \n & \n & \n \\
CoderChrome \cite{Harward10in} & 2010 & \y & \n & \n & \n & \y & \y & \y & \y & \y & \n & \n & \y & \y & \y & \n & \n & \y & \y & \y & \y & \n & \y & \n & \y & \y & \n & \n & \y \\
CodeTalk \cite{Steinert10codetalk} & 2010 & \n & \n & \y & \n & \n & \n & \y & \n & \n & \n & \n & \y & \n & \n & \n & \n & \n & \y & \n & \n & \n & \n & \n & \n & \n & \n & \n & \y \\
Collabode \cite{Goldman11real} & 2011 & \n & \n & \n & \n & \y & \n & \y & \n & \n & \n & \n & \y & \n & \y & \n & \n & \y & \y & \n & \n & \n & \y & \n & \n & \n & \n & \y & \n \\\hline
CoRED \cite{Lautamaeki12cored} & 2012 & \n & \n & \n & \y & \y & \y & \n & \y & \n & \n & \n & \y & \n & \n & \n & \n & \n & \y & \n & \n & \n & \y & \n & \y & \n & \y & \n & \y \\
CostHat \cite{Leitner16modelling} & 2016 & \y & \n & \n & \n & \n & \y & \n & \n & \n & \n & \n & \y & \n & \n & \n & \n & \n & \y & \n & \n & \n & \y & \n & \y & \n & \n & \n & \y \\
CSeR \cite{Jacob10actively} & 2010 & \n & \n & \n & \y & \n & \n & \y & \n & \n & \n & \n & \y & \n & \n & \n & \n & \n & \y & \n & \n & \n & \y & \n & \y & \n & \n & \n & \y \\
CSIclipse \cite{Ohmann15csiclipse} & 2015 & \y & \y & \n & \n & \n & \n & \y & \n & \n & \n & \n & \y & \n & \y & \n & \n & \y & \y & \n & \y & \y & \n & \n & \y & \y & \n & \n & \y \\
CssCoco \cite{Goncharenko16language} & 2016 & \y & \n & \n & \n & \n & \y & \n & \n & \n & \n & \n & \n & \y & \y & \n & \n & \y & \y & \n & \y & \n & \n & \n & \n & \n & \n & \y & \y \\\hline
Debugger Canvas \cite{DeLine12debugger} & 2012 & \n & \y & \n & \n & \n & \n & \n & \y & \n & \n & \y & \y & \y & \y & \y & \n & \y & \n & \y & \y & \n & \n & \n & \y & \n & \n & \n & \y \\
Diver \cite{Myers10using} & 2010 & \n & \y & \n & \n & \n & \y & \n & \n & \n & \n & \n & \y & \n & \n & \n & \n & \y & \n & \n & \n & \y & \n & \n & \n & \n & \n & \y & \y \\
DrScheme \cite{Findler02drscheme} & 2002 & \y & \n & \n & \n & \n & \y & \n & \n & \n & \n & \y & \n & \n & \n & \y & \n & \n & \y & \n & \n & \n & \y & \n & \y & \n & \y & \n & \n \\
DSketch \cite{Cossette10dsketch} & 2010 & \y & \n & \n & \n & \n & \n & \n & \y & \n & \n & \n & \y & \n & \y & \n & \n & \y & \y & \n & \y & \n & \y & \n & \n & \y & \n & \n & \y \\
Dynamic text \cite{Chiba12do} & 2012 & \n & \n & \y & \y & \n & \n & \y & \n & \n & \n & \n & \y & \n & \n & \n & \n & \n & \y & \n & \n & \n & \y & \n & \n & \n & \n & \y & \y \\\hline
Eclipse PTP \cite{Watson06developing} & 2006 & \y & \n & \n & \n & \n & \y & \n & \n & \n & \n & \n & \n & \n & \y & \n & \n & \y & \n & \n & \y & \n & \n & \n & \n & \n & \n & \y & \y \\
EG \cite{Edwards04example} & 2004 & \n & \y & \n & \n & \n & \n & \y & \n & \n & \n & \n & \y & \n & \n & \n & \n & \n & \y & \n & \n & \n & \y & \n & \y & \n & \n & \n & \y \\
eMoose \cite{Dekel09improving} & 2009 & \n & \n & \y & \n & \n & \y & \y & \n & \n & \n & \n & \n & \y & \y & \n & \n & \y & \y & \n & \y & \n & \y & \n & \y & \n & \n & \n & \y \\
EnergyDebugger \cite{Banerjee16debugging} & 2016 & \n & \y & \n & \n & \y & \y & \n & \n & \n & \n & \n & \y & \n & \n & \n & \n & \n & \y & \n & \n & \y & \n & \n & \n & \n & \n & \y & \y \\
Error notifications \cite{Barik14compiler} & 2014 & \y & \n & \n & \n & \n & \n & \n & \n & \n & \n & \y & \n & \n & \n & \y & \n & \n & \y & \n & \n & \n & \y & \n & \n & \n & \n & \y & \n \\\hline
FeatureIDE \cite{Meinicke16featureide} & 2016 & \y & \n & \n & \n & \n & \n & \n & \y & \n & \n & \n & \y & \n & \n & \n & \n & \y & \y & \y & \n & \y & \n & \n & \n & \n & \n & \y & \y \\
Featureous \cite{Olszak12remodularizing} & 2012 & \n & \y & \y & \n & \n & \n & \n & \n & \y & \n & \n & \y & \n & \n & \n & \n & \y & \n & \n & \n & \y & \n & \n & \y & \n & \n & \n & \y \\
FixBugs \cite{Barik16from} & 2016 & \y & \n & \n & \n & \n & \n & \y & \y & \n & \n & \n & \y & \y & \y & \n & \n & \y & \y & \n & \n & \n & \y & \n & \n & \n & \y & \n & \y \\
FLOrIDA \cite{Andam17florida} & 2017 & \y & \n & \n & \n & \n & \n & \n & \y & \n & \n & \n & \y & \n & \n & \n & \n & \n & \y & \n & \y & \n & \n & \n & \n & \n & \n & \y & \n \\
Flower \cite{Smith17flower} & 2017 & \y & \n & \n & \n & \n & \y & \n & \n & \n & \n & \n & \n & \y & \n & \n & \n & \n & \y & \n & \n & \n & \y & \n & \n & \y & \n & \n & \y \\\hline
Fluid views \cite{Desmond06fluid} & 2006 & \y & \n & \n & \n & \n & \y & \n & \n & \n & \y & \n & \y & \y & \n & \n & \y & \n & \y & \n & \n & \n & \y & \n & \n & \n & \y & \n & \y \\
FluidEdt \cite{Ou15interactive} & 2015 & \n & \y & \n & \n & \n & \n & \n & \n & \n & \n & \y & \n & \n & \n & \y & \n & \y & \n & \n & \y & \n & \n & \n & \n & \n & \y & \n & \n \\
Fractional owner. \cite{Mueller15in} & 2015 & \n & \n & \n & \n & \y & \n & \n & \n & \n & \n & \y & \n & \n & \n & \y & \n & \n & \n & \y & \y & \n & \n & \n & \y & \n & \n & \n & \y \\
GhostFactor \cite{Ge14manual} & 2014 & \y & \n & \n & \y & \n & \y & \n & \n & \n & \n & \n & \n & \y & \n & \n & \n & \n & \y & \n & \n & \n & \y & \n & \y & \n & \n & \n & \y \\
Gibber \cite{Roberts14gibber} & 2014 & \n & \y & \n & \n & \n & \n & \n & \n & \n & \n & \y & \n & \n & \n & \y & \n & \n & \y & \n & \n & \n & \n & \y & \n & \n & \n & \y & \n \\\hline
Gilligan \cite{Holmes07task} & 2007 & \y & \n & \y & \n & \n & \n & \y & \n & \n & \n & \n & \y & \n & \y & \n & \n & \y & \y & \n & \y & \n & \y & \n & \n & \n & \n & \y & \y \\
Hermion \cite{Roethlisberger08exploiting} & 2008 & \n & \y & \n & \n & \n & \y & \y & \n & \n & \n & \n & \n & \n & \y & \n & \n & \n & \y & \n & \y & \n & \y & \n & \y & \y & \n & \n & \y \\\hline

\end{tabular}
\end{table}

\begin{table}
\vspace*{-3cm}
\caption{Augmentation tools (continued)}
\label{t:tools2}
\hspace*{-2cm}
\setlength{\tabcolsep}{0.38em}
\renewcommand{\arraystretch}{1.14}
\centering
\footnotesize
\begin{tabular}{|l|l|ccccc|cccccc|ccccc|ccc|cccc|cccc|c|} \hline
\multirow{2}{*}{\textbf{Tool}}
 & \multirow{2}{*}{\textbf{Year}}
 & \multicolumn{5}{c|}{\textbf{Source}}
 & \multicolumn{6}{c|}{\textbf{Type}}
 & \multicolumn{5}{c|}{\textbf{Visualization}}
 & \multicolumn{3}{c|}{\textbf{Location}}
 & \multicolumn{4}{c|}{\textbf{Target}}
 & \multicolumn{4}{c|}{\textbf{Interaction}}
 & \multicolumn{1}{c|}{\textbf{IDE}} \\ \cline{3-30}

~ & ~ & \h{code} & \h{runtime} & \h{human} & \h{interaction} & \h{collaboration} & \h{boolean} & \h{fixed enum.} & \h{variable enum.} & \h{number} & \h{string} & \h{object} & \h{color} & \h{decoration} & \h{icon} & \h{graphics} & \h{text} & \h{left} & \h{in code} & \h{right} & \h{line} & \h{line range} & \h{character range} & \h{file} & \h{popover} & \h{navigate} & \h{change} & \h{none/unknown} & \h{existing} \\ \hline

Historef \cite{Hayashi15historef} & 2015 & \n & \n & \y & \y & \n & \n & \y & \n & \n & \n & \n & \y & \n & \n & \n & \n & \n & \y & \n & \n & \n & \y & \n & \n & \n & \n & \y & \y \\
HyperSource \cite{Hartmann11hypersource} & 2011 & \y & \n & \n & \y & \n & \n & \y & \n & \n & \n & \n & \n & \y & \y & \n & \n & \y & \n & \n & \y & \n & \n & \n & \y & \n & \n & \n & \y \\
I3 \cite{Beck15rethinking} & 2015 & \y & \n & \n & \n & \y & \n & \y & \n & \n & \n & \y & \y & \n & \n & \y & \n & \n & \y & \n & \n & \n & \y & \n & \y & \n & \n & \n & \y \\
IDE sparklines \cite{Beck13visual} & 2013 & \n & \y & \n & \n & \n & \n & \n & \n & \y & \n & \n & \n & \n & \n & \y & \n & \n & \y & \n & \y & \n & \n & \n & \y & \n & \n & \n & \y \\
IDEOL \cite{Dang14educo} & 2014 & \n & \n & \n & \n & \y & \n & \n & \y & \n & \n & \n & \y & \n & \n & \n & \y & \n & \y & \y & \n & \n & \y & \n & \n & \n & \n & \y & \n \\\hline
iMaus \cite{Kawrykow09improving} & 2009 & \y & \n & \n & \n & \n & \y & \n & \n & \n & \n & \n & \n & \y & \y & \n & \n & \y & \y & \n & \y & \n & \y & \n & \n & \y & \n & \n & \y \\
Impromptu HUD \cite{Swift13visual} & 2013 & \n & \y & \n & \n & \n & \n & \n & \n & \y & \n & \n & \n & \n & \n & \y & \y & \n & \y & \n & \y & \n & \n & \y & \n & \n & \n & \y & \y \\
In situ profiler \cite{Beck13in} & 2013 & \n & \y & \n & \n & \n & \n & \y & \y & \y & \n & \n & \n & \n & \n & \y & \n & \n & \y & \n & \y & \n & \n & \n & \y & \n & \n & \n & \y \\
inCode \cite{Ganea17continuous} & 2017 & \y & \n & \n & \n & \n & \y & \n & \n & \n & \n & \n & \n & \n & \y & \n & \n & \y & \n & \n & \y & \n & \n & \n & \y & \n & \n & \n & \y \\
IVCon \cite{Saigal09inline} & 2009 & \n & \n & \y & \n & \n & \n & \n & \y & \n & \n & \n & \y & \n & \n & \n & \n & \n & \y & \n & \n & \n & \y & \n & \y & \n & \n & \n & \n \\\hline
iXj \cite{Boshernitsan07aligning} & 2007 & \y & \n & \n & \n & \n & \y & \n & \n & \n & \n & \n & \y & \n & \n & \n & \n & \y & \y & \n & \n & \n & \y & \n & \n & \y & \n & \n & \y \\
Jazz Band \cite{Hupfer04introducing} & 2004 & \n & \n & \n & \n & \y & \n & \y & \n & \n & \n & \n & \y & \n & \y & \n & \n & \y & \n & \n & \n & \y & \n & \n & \y & \y & \n & \n & \y \\
JDeodorant \cite{Tsantalis08jdeodorant} & 2008 & \y & \n & \n & \n & \n & \y & \n & \n & \n & \n & \n & \y & \n & \n & \n & \n & \n & \y & \n & \n & \n & \y & \n & \n & \n & \n & \y & \y \\
jGRASP \cite{CrossII11combining} & 2011 & \y & \n & \n & \n & \n & \n & \n & \n & \n & \n & \y & \n & \n & \n & \y & \n & \n & \y & \n & \n & \n & \n & \y & \n & \n & \y & \n & \n \\
Jigsaw \cite{Cottrell08semi} & 2008 & \y & \n & \n & \y & \n & \n & \y & \n & \n & \n & \n & \y & \n & \n & \n & \n & \n & \y & \n & \n & \n & \y & \n & \n & \n & \n & \y & \y \\\hline
JScoper \cite{Ferrari05jscoper} & 2005 & \y & \n & \n & \n & \n & \y & \n & \n & \n & \n & \n & \n & \n & \y & \n & \n & \y & \n & \n & \y & \n & \n & \n & \n & \y & \n & \n & \y \\
Live coding IDE \cite{Kramer14how} & 2014 & \n & \y & \n & \n & \n & \n & \n & \n & \y & \y & \n & \n & \n & \n & \n & \y & \n & \n & \y & \y & \n & \n & \n & \n & \n & \y & \n & \y \\
Moonstone \cite{Kistner17moonstone} & 2017 & \y & \n & \n & \n & \n & \y & \y & \n & \n & \y & \n & \y & \y & \y & \n & \y & \y & \y & \n & \y & \y & \y & \n & \y & \n & \n & \n & \y \\
Morpheus \cite{Eyl17traceability} & 2017 & \y & \n & \n & \y & \n & \n & \y & \n & \n & \y & \n & \y & \y & \y & \n & \y & \y & \y & \n & \y & \y & \y & \n & \y & \y & \y & \n & \y \\
NLP Eclipse \cite{Witte11intelligent} & 2011 & \y & \n & \n & \n & \n & \y & \n & \n & \n & \n & \n & \n & \n & \y & \n & \n & \y & \n & \n & \n & \n & \y & \n & \y & \n & \n & \n & \y \\\hline
Omnicode \cite{Kang17omnicode} & 2017 & \y & \n & \y & \y & \n & \y & \n & \n & \n & \n & \n & \y & \n & \n & \n & \n & \y & \y & \n & \y & \n & \y & \n & \y & \y & \n & \n & \n \\
PerformanceHat \cite{Cito15runtime} & 2015 & \n & \y & \n & \n & \n & \y & \n & \n & \n & \n & \n & \y & \n & \y & \n & \n & \y & \y & \n & \n & \n & \y & \n & \y & \n & \n & \n & \y \\
Pollicino \cite{Guzzi11collective} & 2011 & \n & \n & \y & \n & \n & \y & \n & \n & \n & \n & \n & \y & \n & \y & \n & \n & \y & \y & \n & \y & \n & \n & \n & \n & \y & \n & \n & \y \\
PyLighter \cite{Boland09introducing} & 2009 & \n & \y & \n & \n & \n & \n & \n & \n & \y & \n & \n & \y & \n & \n & \n & \n & \n & \y & \n & \y & \n & \n & \n & \n & \n & \n & \y & \n \\
Rationalizer \cite{Bradley11supporting} & 2011 & \n & \n & \n & \n & \y & \n & \y & \y & \y & \y & \n & \y & \n & \y & \n & \y & \n & \n & \y & \y & \n & \n & \n & \y & \y & \n & \n & \y \\\hline
Refactoring Ann. \cite{Murphy-Hill08breaking} & 2008 & \y & \n & \n & \n & \n & \n & \n & \n & \n & \n & \y & \n & \n & \n & \y & \n & \n & \y & \n & \n & \n & \y & \n & \n & \n & \n & \y & \y \\
RegViz \cite{Beck14regviz} & 2014 & \y & \n & \n & \n & \n & \n & \y & \n & \y & \n & \n & \y & \y & \n & \n & \y & \n & \y & \n & \n & \n & \y & \n & \n & \n & \y & \n & \n \\
Remail \cite{Bacchelli11miler} & 2011 & \n & \n & \n & \n & \y & \y & \n & \n & \n & \n & \n & \n & \n & \y & \n & \n & \y & \n & \n & \y & \n & \n & \n & \y & \n & \n & \n & \y \\
Saros \cite{Salinger10saros} & 2010 & \n & \n & \n & \n & \y & \n & \n & \y & \n & \n & \n & \y & \n & \n & \n & \n & \y & \y & \n & \n & \y & \y & \n & \n & \y & \n & \n & \y \\
SE-Editor \cite{Schugerl09beyond} & 2009 & \n & \n & \y & \n & \n & \n & \n & \n & \n & \n & \y & \n & \n & \n & \y & \n & \n & \y & \n & \n & \y & \n & \n & \n & \n & \n & \y & \y \\\hline
Senseo \cite{Roethlisberger12exploiting} & 2012 & \n & \y & \n & \n & \n & \n & \y & \n & \n & \n & \n & \y & \n & \n & \y & \n & \y & \n & \y & \y & \n & \n & \n & \n & \n & \n & \y & \y \\
Shared-code editor \cite{Lee17exploring} & 2017 & \y & \n & \y & \y & \y & \y & \n & \n & \n & \y & \n & \n & \y & \n & \n & \y & \n & \n & \n & \n & \y & \n & \n & \n & \n & \n & \y & \n \\
SimEclipse \cite{Uddin15towards} & 2015 & \y & \n & \n & \n & \n & \y & \n & \n & \n & \n & \n & \y & \n & \y & \n & \n & \y & \y & \n & \y & \y & \n & \n & \y & \y & \y & \n & \y \\
SketchLink \cite{Baltes14linking} & 2014 & \y & \n & \n & \n & \n & \y & \n & \n & \n & \n & \n & \n & \n & \y & \n & \n & \n & \y & \n & \y & \n & \n & \n & \y & \y & \y & \n & \y \\
Soot-Eclipse \cite{Lhotak04integrating} & 2004 & \y & \n & \n & \n & \n & \y & \n & \n & \n & \n & \n & \y & \n & \y & \n & \n & \y & \y & \n & \y & \n & \y & \n & \y & \y & \n & \n & \y \\\hline
Soul \cite{DeRoover11soul} & 2011 & \y & \n & \n & \n & \n & \y & \n & \n & \n & \n & \n & \n & \n & \y & \n & \n & \y & \n & \n & \y & \n & \n & \n & \y & \n & \n & \n & \y \\
SourcererCC-I \cite{Saini16sourcerercc} & 2016 & \y & \n & \n & \n & \n & \n & \y & \n & \n & \n & \n & \n & \n & \y & \n & \n & \y & \n & \n & \y & \n & \n & \n & \n & \n & \n & \y & \y \\
Spotlight \cite{Revelle05understanding} & 2005 & \n & \n & \y & \n & \n & \n & \n & \y & \n & \n & \n & \y & \y & \n & \n & \n & \y & \y & \n & \n & \y & \n & \n & \n & \n & \n & \y & \y \\
Stacksplorer \cite{Karrer11stacksplorer} & 2011 & \y & \n & \n & \n & \n & \y & \y & \y & \n & \y & \n & \y & \y & \y & \n & \y & \y & \y & \y & \n & \y & \y & \n & \n & \y & \y & \n & \y \\
Stench Blossom \cite{Murphy-Hill10interactive} & 2010 & \y & \n & \n & \n & \n & \n & \n & \n & \y & \n & \n & \n & \n & \n & \y & \n & \n & \n & \y & \n & \n & \n & \y & \y & \n & \n & \n & \y \\\hline
Syde \cite{Hattori10syde} & 2010 & \n & \n & \n & \n & \y & \n & \y & \n & \n & \n & \n & \y & \n & \n & \n & \n & \y & \n & \n & \n & \y & \n & \n & \y & \n & \n & \n & \y \\
TagSEA \cite{Storey09how} & 2009 & \n & \n & \y & \n & \n & \y & \n & \n & \n & \n & \n & \n & \n & \y & \n & \n & \y & \n & \n & \y & \n & \n & \n & \n & \y & \n & \n & \y \\
TexMo \cite{Pfeiffer12texmo} & 2012 & \y & \n & \n & \n & \n & \n & \y & \n & \n & \n & \n & \y & \n & \y & \n & \n & \y & \y & \n & \y & \n & \y & \n & \y & \n & \n & \n & \y \\
Theseus \cite{Lieber14addressing} & 2014 & \n & \y & \n & \n & \n & \n & \y & \n & \y & \y & \n & \n & \n & \y & \n & \y & \y & \n & \n & \y & \n & \y & \n & \n & \y & \y & \n & \y \\
Traces view \cite{Alsallakh12visual} & 2012 & \n & \y & \n & \n & \n & \y & \n & \n & \n & \n & \n & \y & \n & \n & \n & \n & \n & \y & \n & \y & \n & \n & \n & \n & \n & \n & \y & \y \\\hline
vLens \cite{Li13calculating} & 2013 & \y & \y & \n & \n & \n & \n & \n & \n & \y & \n & \n & \y & \n & \n & \n & \n & \n & \n & \n & \n & \y & \n & \n & \n & \n & \n & \y & \y \\
wIDE \cite{Murolo17improved} & 2017 & \y & \n & \n & \n & \n & \n & \y & \n & \n & \n & \n & \y & \n & \n & \n & \n & \n & \y & \n & \n & \n & \y & \n & \n & \n & \n & \y & \y \\
xLineMapper \cite{Zheng17mapping} & 2017 & \n & \n & \y & \n & \n & \y & \n & \y & \n & \n & \n & \y & \n & \y & \n & \n & \y & \y & \n & \y & \n & \y & \n & \n & \n & \n & \y & \y \\
XSS marker \cite{Bathia11assisting} & 2011 & \y & \n & \n & \n & \n & \y & \n & \n & \n & \n & \n & \n & \y & \y & \n & \n & \y & \y & \n & \y & \n & \y & \n & \y & \n & \n & \n & \y \\
YinYang \cite{McDirmid13usable} & 2013 & \n & \y & \n & \n & \n & \n & \n & \n & \n & \y & \n & \n & \n & \n & \n & \y & \n & \y & \n & \y & \n & \n & \n & \n & \y & \n & \n & \n \\\hline
Zelda \cite{Ratanotayanon09supporting} & 2009 & \y & \n & \y & \n & \n & \y & \n & \n & \n & \n & \n & \n & \n & \y & \n & \n & \y & \n & \n & \y & \n & \n & \n & \n & \n & \n & \y & \y \\\hline

\end{tabular}
\end{table}

If the tool contains an augmentation fulfilling the given attribute, this is denoted by a filled square (\y). Otherwise, an empty square (\n) is displayed.

Note that the ``IDE'' dimension has only one attribute displayed since it has two mutually exclusive attributes. Also note that one tool can contain multiple related or unrelated augmentations (e.g. an icon and color highlighting) or one augmentation fulfilling multiple attributes (e.g., an icon offering both popover and navigation possibilities).

\begin{figure}
\centering
\includegraphics[scale=0.6]{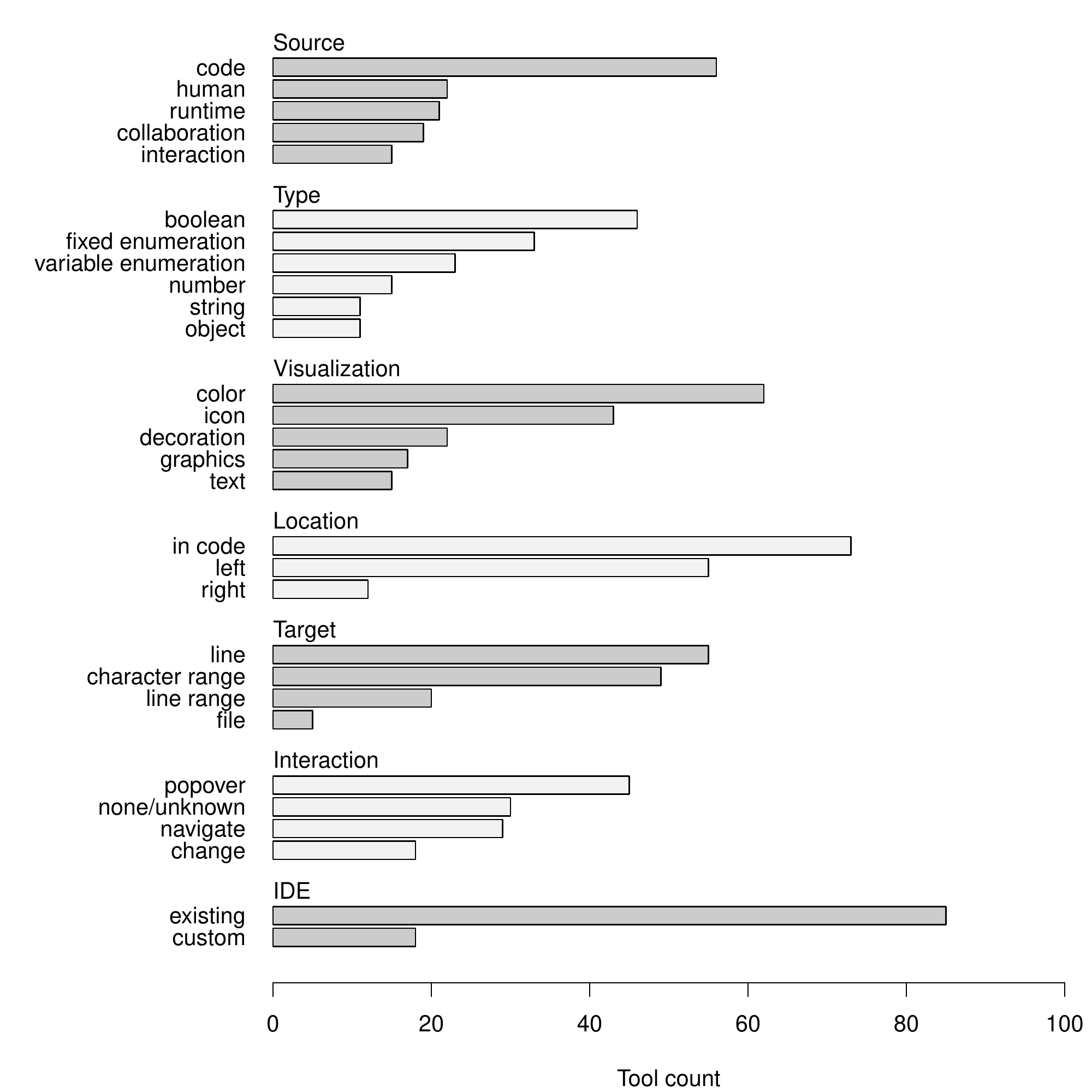}
\caption{The numbers of tools fulfilling the given attributes of the dimensions} \label{f:attrib}
\end{figure}

In Figure~\ref{f:attrib}, there is a chart displaying the numbers of reviewed tools fulfilling each attribute of a particular dimension. In each dimension, the attributes are ordered by frequency.

It can be seen from the chart that the most popular source of data for augmentation is the code itself. This is understandable because source code augmentation was traditionally used by static analysis tools. As can be expected, simple visualization techniques like color or icon and simple data (boolean, fixed enumeration) were used more often than complex visualizations of complex data that are much harder to design and implement. Graphical visualizations are also rarely used by the industrial IDEs discussed in Section \ref{s:ides}, although they actually implement augmentations from all attributes of our taxonomy.

Most of the visualizations were displayed directly in code or in the left margin. The right margin was used by only a few tools. This may be related to the fact that majority of the tools were based on existing industrial IDEs and these environments provide direct support for highlighting parts of code and displaying markers in the left margin. Some of these IDEs can also automatically display an overview of all left-margin icons in the right margin, but since this is their built-in feature, it is not shown in the tables.

The least represented attribute in our review is the \emph{file} target. Integration of some information directly in the code editor is usually most useful if it is related to some specific parts of the code.

\begin{figure}
\centering
\includegraphics[scale=0.6]{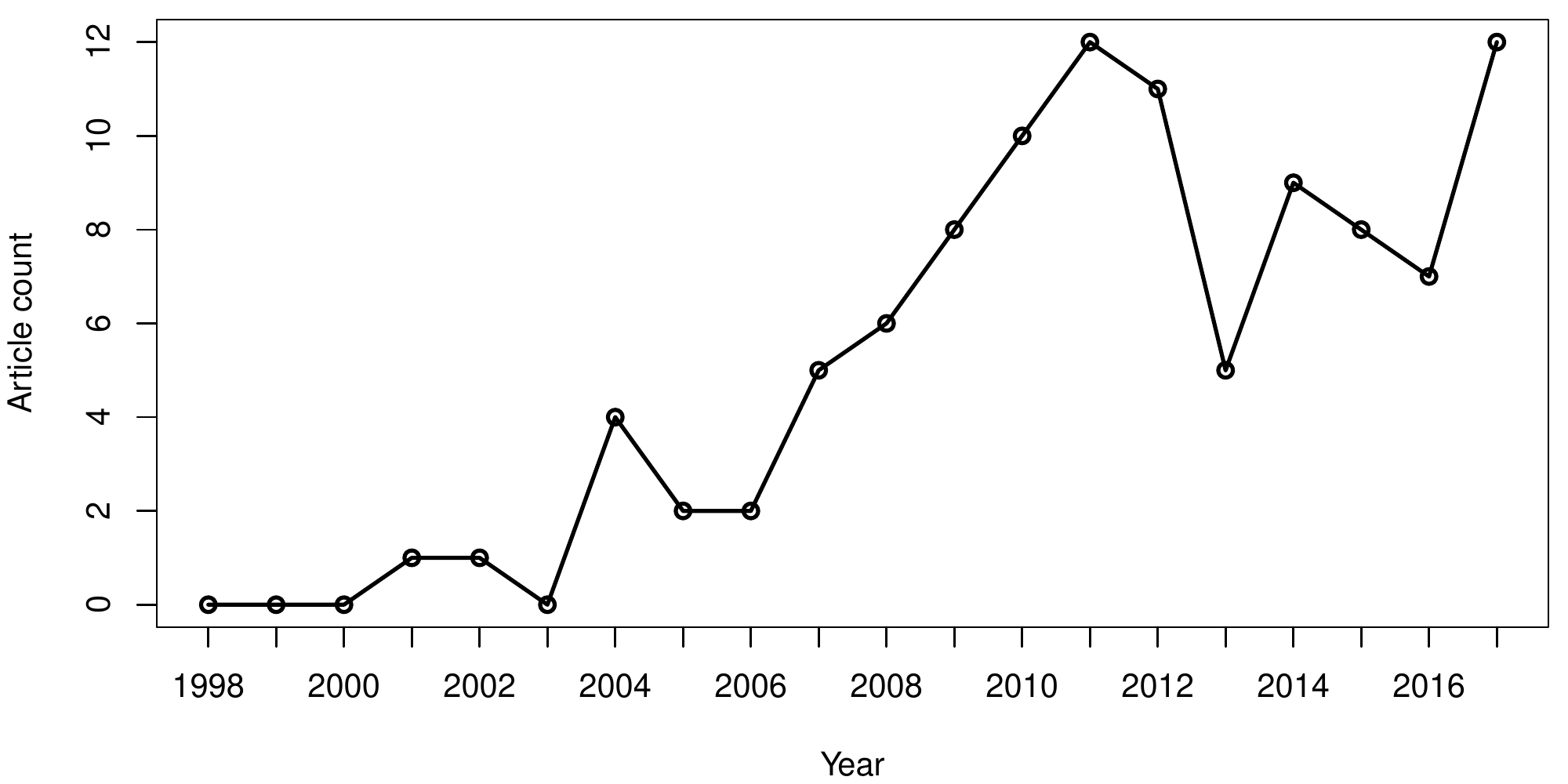}
\caption{A distribution of the publications over years} \label{f:years}
\end{figure}

In Figure~\ref{f:years}, we can see a distribution of the 103 reviewed publications over years 1998--2017. After a slow start, the number of articles per year tends to rise from 2005 to 2011 and then begins to oscillate. The sudden growth starting in the mid-2000s can be possibly attributed to the ascending popularity of the IDEs, particularly Eclipse. Note, however, that the numbers are only indicative since some articles might have been missed during the systematic review process.

We have also analyzed venues where the selected papers were published. Most of the papers (94) were published at conferences, while only 9 were published in journals. List of the top venues is provided in Table \ref{t:venues}. It contains all venues with more than one publication in our study. Clearly, the most popular venue is the International Conference on Software Engineering (ICSE) with 18 publications.

\begin{table}
  \caption{Top venues with count of publications used in the study}
  \label{t:venues}
  \small
  \renewcommand{\arraystretch}{1.3}
  \centering
  \begin{tabular}{p{10.8cm}r}
    \textbf{Venue} & \textbf{Count} \\
    \hline
    International Conference on Software Engineering (ICSE) & 18 \\
    International Conference on Program Comprehension (ICPC) & 7 \\
    Conference on Human Factors in Computing Systems (CHI) & 6 \\
    International symposium on Foundations of software engineering (FSE) & 5 \\
    Symposium on Visual Languages and Human-Centric Computing (VL/HCC) & 5 \\
    Symposium on User Interface Software and Technology (UIST) & 4 \\
    Conference on Computer Supported Cooperative Work (CSCW) & 2 \\
    Eclipse Technology eXchange (ETX) & 2 \\
    European Conference on Software Maintenance and Reengineering (CSMR) & 2 \\
    International Conference on Automated Software Engineering (ASE) & 2 \\
    International Symposium on New Ideas, New Paradigms, and Reflections on Programming and Software (Onward!) & 2 \\
    IEEE Transactions on Software Engineering & 2 \\
    Science of Computer Programming & 2 \\
    \hline
  \end{tabular}
\end{table}

\section{Conclusion}

We presented a systematic mapping study about visual augmentation of source code editors. Using keyword and references search combined with manual filtering, we constructed a list containing 103 relevant tools described in research articles. A taxonomy representing distinct and similar characteristics of these tools was constructed. It contains seven dimensions: source, type, visualization, location, target, interaction and IDE. Each tool was characterized according to this taxonomy, producing a table with article references. The assignment of the features (dimensions' attributes) to the tools provides a brief overview of their properties and can be used for comparison.

Our main contributions are:
\begin{itemize}
\item the definition of the term ``source code editor augmentation'',
\item a taxonomy of source code editor augmentation features and
\item a categorized list of augmentation tools with references.
\end{itemize}
This article can be a useful resource for researchers aiming to gain an overview of this area or finding a particular example of given augmentations. Furthermore, we provide multiple directions for future work, which we will now describe.

Since we found more than 100 tools augmenting the source code editor area, there is a need to think about filtering possibilities. Even though some augmentations have only a limited lifetime, the IDE may easily become visually cluttered. An interesting question is also how to resolve conflicts when displaying visually overlapping augmentations.

The underlying data or source code often change after the augmentation is initially displayed. Invalidation and possible recalculation of augmentations is another important issue, seldom discussed in the reviewed articles.

There are many papers mentioning source code editor augmentation; nevertheless, it is often only a marginal topic. Although in some articles, augmentation is a central topic (such as in our recent paper about variable values in the code \cite{Sulir18augmenting}), they often miss empirical evaluation. On the other hand, while many articles include empirical evaluation, the visual code augmentation itself is rarely the main object of the studies. One of the notable exceptions is a recent runtime state visualization approach by Hoffswell et al. \cite{Hoffswell18augmenting}, accompanied by a thorough empirical study. Further comparisons of the usability of separate views, in-code augmentations and their variations are definitely welcome.

\section*{Acknowledgment}

This work was supported by project KEGA 047TUKE-4/2016 Integrating software processes into the teaching of programming.

\section*{References}

\bibliography{jvlc}

\end{document}